 \documentstyle[preprint,prl,aps,epsfig]{revtex}
\begin{document}

\draft                    % This command makes PACS numbers print

%---------------------------------------------------------------------

\title{
         \vskip -0.5cm
         \hfill\hfil{\rm\normalsize Printed on \today}\\
         Giant magneto-conductance
         in twisted carbon nanotubes }

\author{
         Steven W.D.~Bailey$^{1}$,
         David Tom\'anek,$^{2}$
        Young-Kyun Kwon,$^{2,}$\cite{KwonUCB} and
         Colin J.~Lambert$^{1}$
        }

\address{$^{1}$ Department of Physics, Lancaster University,
         Lancaster, LA1 4YB, UK}

\address{$^{2}$ Department of Physics and Astronomy, and
         Center for Fundamental Materials Research, \\
         Michigan State University,
         East Lansing, Michigan 48824-1116}

\date{Received 22 January 2001 }

\maketitle

\begin{abstract}
Using the Landauer-B\"uttiker formalism, we calculate the effect
of structural twist on electron transport in conducting carbon
nanotubes. We demonstrate that even a localized region of twist
scatters the propagating $\pi$ electrons and induces the opening
of a (pseudo-) gap near the Fermi level. The subsequent
conductance reduction may be compensated by an applied axial
magnetic field, leading to a twist-induced, giant positive
magneto-conductance in clean armchair nanotubes.
\end{abstract}

%---------------------------------------------------------------------
Carbon nanotubes \cite{{Iijima91},{Dresselhaus96},{Saito98}}
exhibit a range of unusual electronic properties  associated with
the morphology of these quasi-1D structures. Early one-electron
theories successfully associated metallic or semiconducting
behaviour with the chiral vector that characterizes a given
nanotube \cite{{Mintmire92},{RSaito92},{Oshiyama92}}. Further
studies addressed the effect of atomic-level impurities
\cite{Roche99,Choi00,Roland00,Anantram98} and inter-tube
interactions \cite{{PRLxDWT},{Louie98},{PRLxNTR},{Sanvito00}} on
electrical conductance, or magneto-transport
\cite{Roche99,AB59,Saito92,Tian94,Lin95}. In parallel with the
development of such one-electron theories, intensive studies of
electron-electron correlations have been undertaken. Indeed,
nonlinear current-voltage ($I-V$) characteristics have recently
been observed \cite{Brockrath99}, which are reminiscent of
Luttinger liquid behaviour. An intriguing question remains
however, namely whether other effects may augment or even dominate
such nonlinearities in the transport properties of nanotubes.

In this letter we predict that scattering of electrons from
twistons may give rise to strongly non-linear $I-V$
characteristics. Twistons \cite{Kane97,Rochefort99,PRLxOM},
associated with regions of axial twist in otherwise perfect
nanotubes, are intrinsic defects that are frozen into nanotube
bundles during their synthesis and hence cannot be ignored when
discussing electron transport. Unlike ideal straight and
defect-free nanotubes, which have been shown to exhibit
conventional magneto-resistive behavior \cite{Maksimenko00}, we
find that nanotubes containing twistons may behave in a very
different way. In twisted tubes, we predict the occurrence of a
giant positive magneto-conductance, an unexpected effect that by
far exceeds the positive magneto-conductance associated with weak
localisation in disordered tubes \cite{Langer94}.

\begin{figure}[ht]

\centering
     \begin{minipage}[c]{0.5\textwidth}
         \hspace*{-0.08\columnwidth}
         {\raisebox{0.0\columnwidth}{\large\bf\textsf {(a)}}}

         \epsfxsize=0.55\columnwidth
         \epsffile{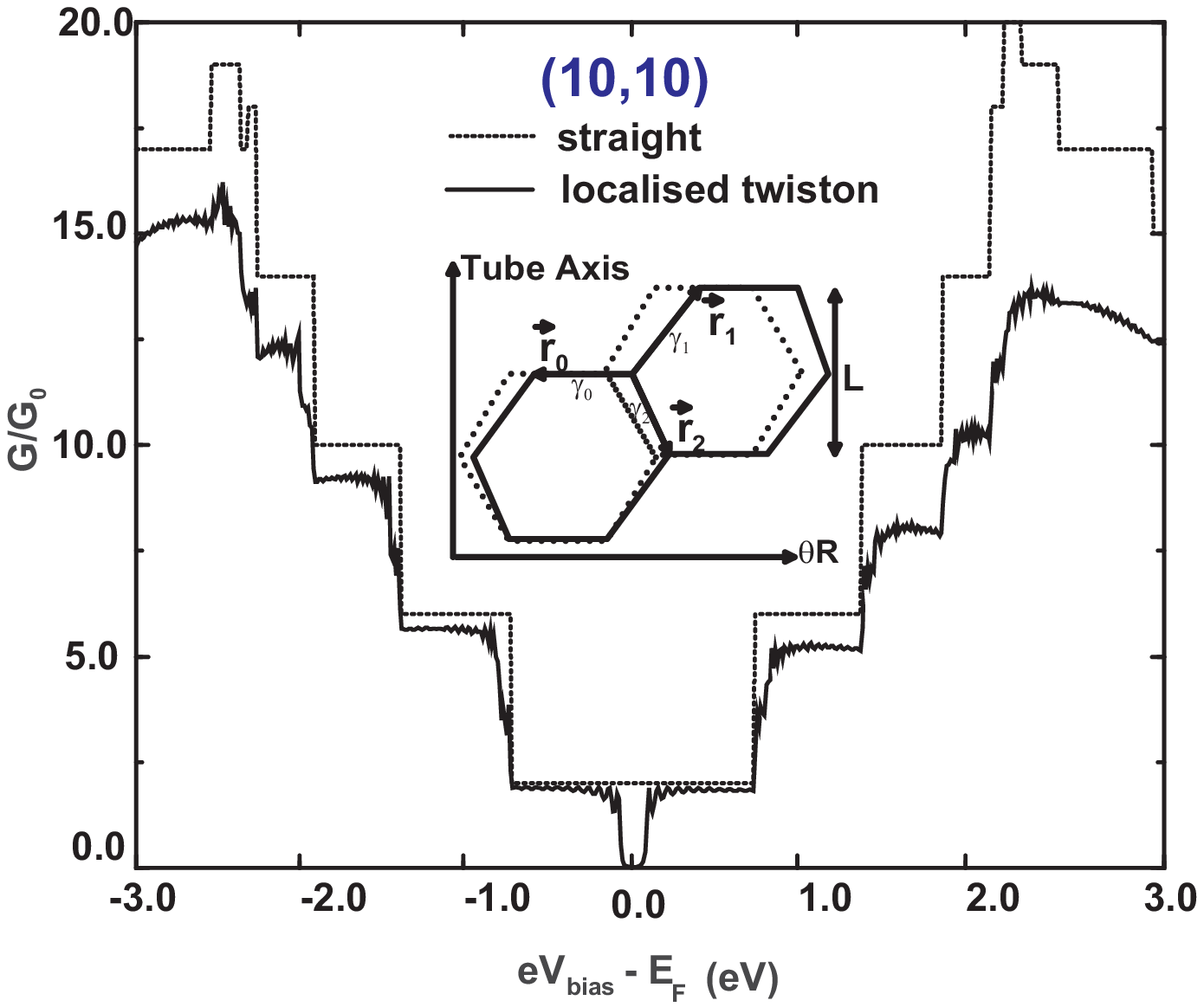}
         \end{minipage}

      \begin{minipage}[c]{0.5\textwidth}

         \hspace*{-0.08\columnwidth}
         {\raisebox{0.5\columnwidth}{\large\bf\textsf {(b)}}}
     {\hspace*{-0.08\columnwidth}
         \epsfxsize=0.65\columnwidth
         \epsffile{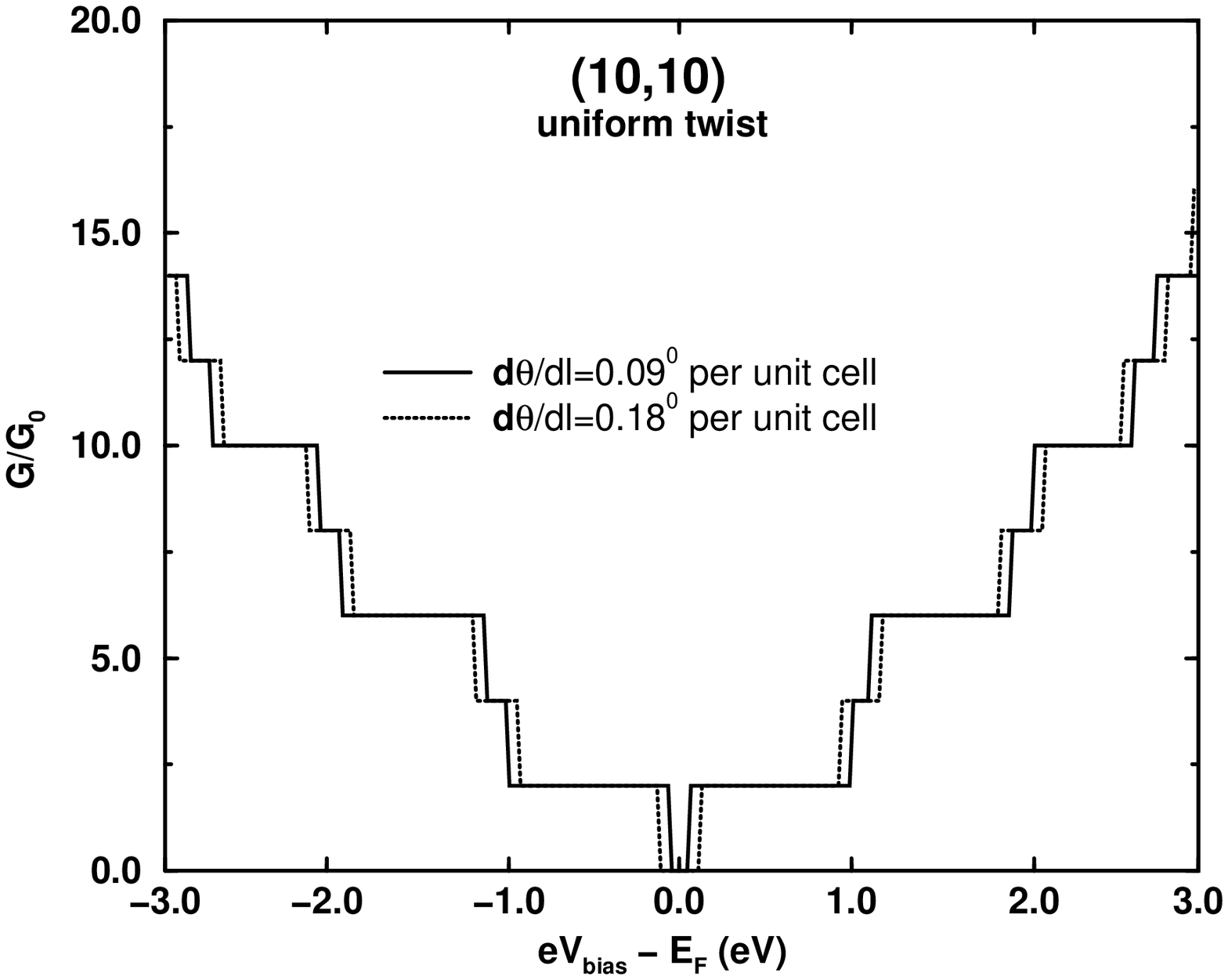}}
          \end{minipage}

    \begin{minipage}[c]{0.5\textwidth}
         \hspace*{-0.08\columnwidth}
         {\raisebox{0.0\columnwidth}{\large\bf\textsf {(c)}}}

         \epsfxsize=0.6\columnwidth
         \epsffile{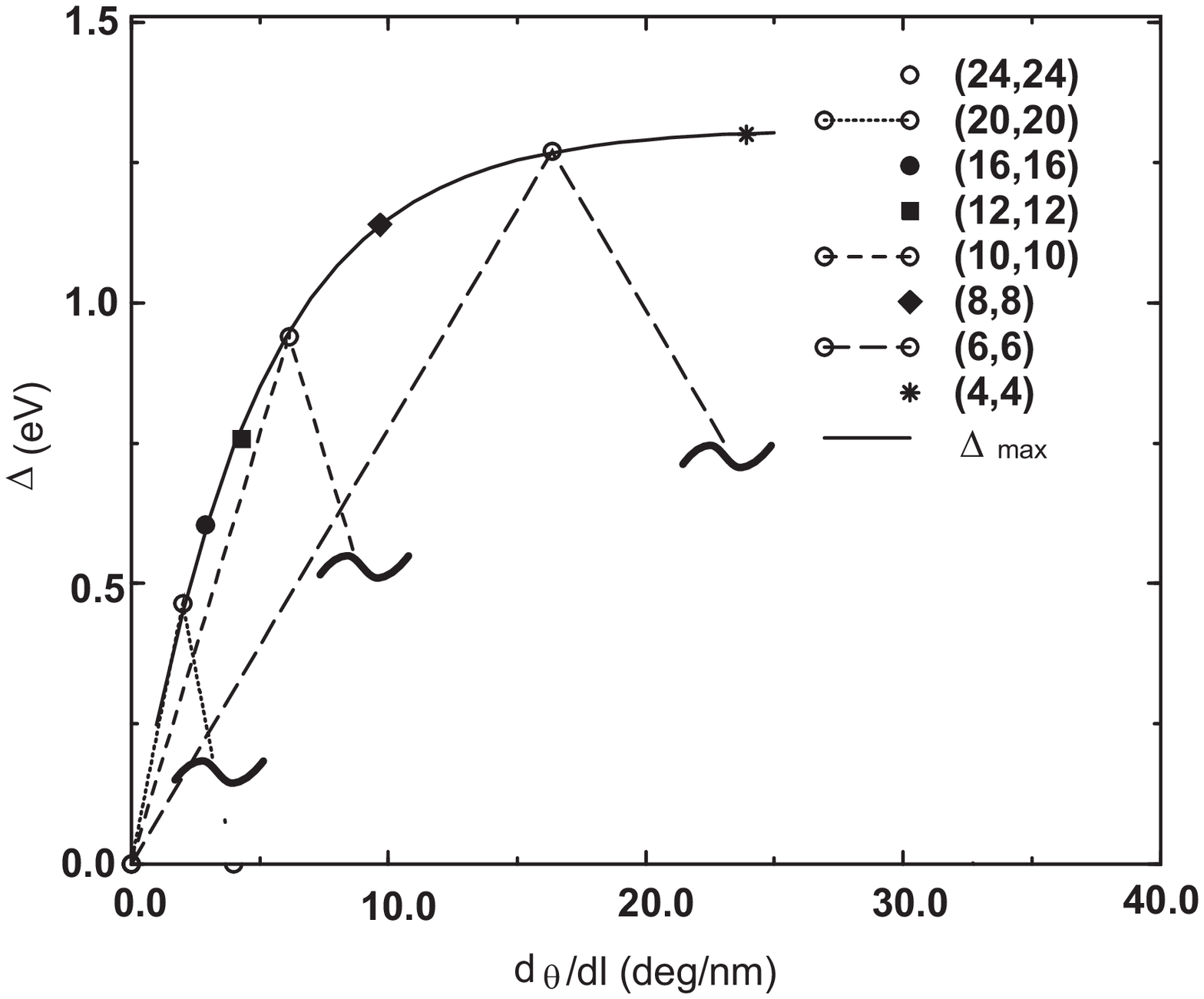}
          \end{minipage}

\vspace*{0.05\columnwidth}%
\caption{ Differential conductance $G$
of an infinite $(10,10)$ carbon nanotube as a function of the
applied bias voltage $V_{\rm bias}$. (a) Results for a perfectly
straight tube (dotted line) are compared to those for a
$\Delta\theta=36^\circ$ finite twiston extending over 100 unit
cells in the axial direction. For a tube with radius $R$, a
portion of the unit cell of length $L$ is shown in the inset.
$G_0={2e^2}/{h}$ is the conductance quantum. (b) Differential
conductance of the $(10,10)$ nanotube subject to an infinitely
long uniform twist, for different values of the twist per unit
length $d\theta/dl$. (c) Dependence of the conductance gap
$\Delta$ on the  infinitely long twist $d\theta/dl$ for various
$(n,n)$ nanotubes. } \label{Fig1}

\end{figure}

To compute the effect of a finite scattering region on transport
in an otherwise perfect $(n,n)$ armchair nanotube, we use a
parameterized four-state ($s, p_x, p_y, p_z$) Hamiltonian, based
on a global fit to Density Functional results for graphite,
diamond and C$_2$ as a function of the lattice parameter
\cite{Tomanek91}. A finite twiston is treated as a scattering
region connecting two semi-infinite $(n,n)$ nanotubes. A recursive
Greens function formalism is used to evaluate the transmission
matrix {\bf t}, describing the scattering of electrons of energy
$E$ from one end of the semi-infinite nanotube to the other
\cite{Sanvito99}. The differential electrical conductance at bias
voltage $V_{\rm bias}$ is related to scattering properties at
energy $E=E_{\rm F}-eV_{\rm bias}$ by the Landauer formula
$G=G_0{\rm Tr}\left\{\mathbf{t^{\dagger}t}\right\}$, where
$G_0={2e^2}/{h}$ is the conductance quantum. A twiston is formed
by introducing a small angular distortion between neighboring
axial slices within the scattering region. This is shown
schematically in the inset of Fig.~\ref{Fig1}(a), which displays
the local shear distortion within the unit cell of length $L$,
containing two axial slices. The perturbation to the Hamiltonian
matrix enters through the scaling of the nearest-neighbor hopping
integrals that follow the changes in C-C bonds {\bf r$_1$} and
{\bf r$_2$}.

Fig.~\ref{Fig1} shows the differential conductance for a straight
and twisted $(10,10)$ nanotube. In the straight nanotube of
Fig.~\ref{Fig1}(a), the non-degenerate bands, which cross near
$E_{\rm F}$, open up two conductance channels at this energy.
Additional conduction channels follow as an increasing number of
subbands appear near $E_{\rm F}-eV_{\rm bias}$ at higher bias
voltages. In the presence of a localized twiston, the differential
conductance is suppressed at all energies, in particular near the
Fermi level. This is shown in Fig.~\ref{Fig1}(a) for a
finite-length twiston, with a total twist of $36^\circ$ extending
over $100$ unit cells or ${\approx}24.6$~nm. The occurrence of a
conductance gap near $E_{\rm F}$ suggests that electron scattering
by a finite twiston can be viewed as a tunneling phenomenon.

For comparison with Fig.~\ref{Fig1}(a), Fig.~\ref{Fig1}(b) shows
results for a nanotube subject to an infinitely long uniform twist
$d\theta/dl$. This shows that that a localized or infinitely long
twist opens up a conductance
% DT 26nov01
% pseudo-gap or
gap $\Delta$ near $E_{\rm F}$. The predicted gap in the presence
of infinitely long twists is shown in Fig.~\ref{Fig1}(c) and
agrees with results of Refs. \cite{Kane97} and \cite{Rochefort99}.

In the following, we discuss the dependence of the conductance gap
on the twist distortion $d\theta/dl$, the spatial extent of the
twiston, and an axially applied magnetic field. As shown in
Fig.~\ref{Fig1}(c), we find that for a range of $(n,n)$ tubes
subject to an infinitely long twist, the magnitude of the
conductance gap $\Delta$ increases linearly with increasing twist
distortion to its maximum value $\Delta_{\max}$, and decreases
thereafter. For an $(n,n)$ nanotube, we find the maximum value of
the conductance gap $\Delta_{\max}$ to be achieved at an ``optimum
tube twist'' value $(d\theta/dl)_0\approx A n^{-2}$ (with
$A=620^{\circ}$/nm), which depends on the chiral index $n$. The
dependence of $\Delta_{\max}$ on the optimum tube twist
$(d\theta/dl)_0$ corresponds to the envelope function in
Fig.~\ref{Fig1}(c), and is well approximated by
$\Delta_{\max}\approx\Delta(d\theta/dl)_0\approx
1.30$~eV~$(1-e^{-0.21(d\theta/dl)_0})$, with $(d\theta/dl)_0$ in
$^{\circ}/$nm units.

\begin{figure}[ht]

\centering
 \begin{minipage}[c]{0.5\textwidth}
         \hspace*{0.04\columnwidth}
         {\raisebox{0.56\columnwidth}{\Large\bf\textsf {(a)}}}
         \hspace*{-0.09\columnwidth}
         \epsfxsize=0.8\columnwidth
         \epsffile{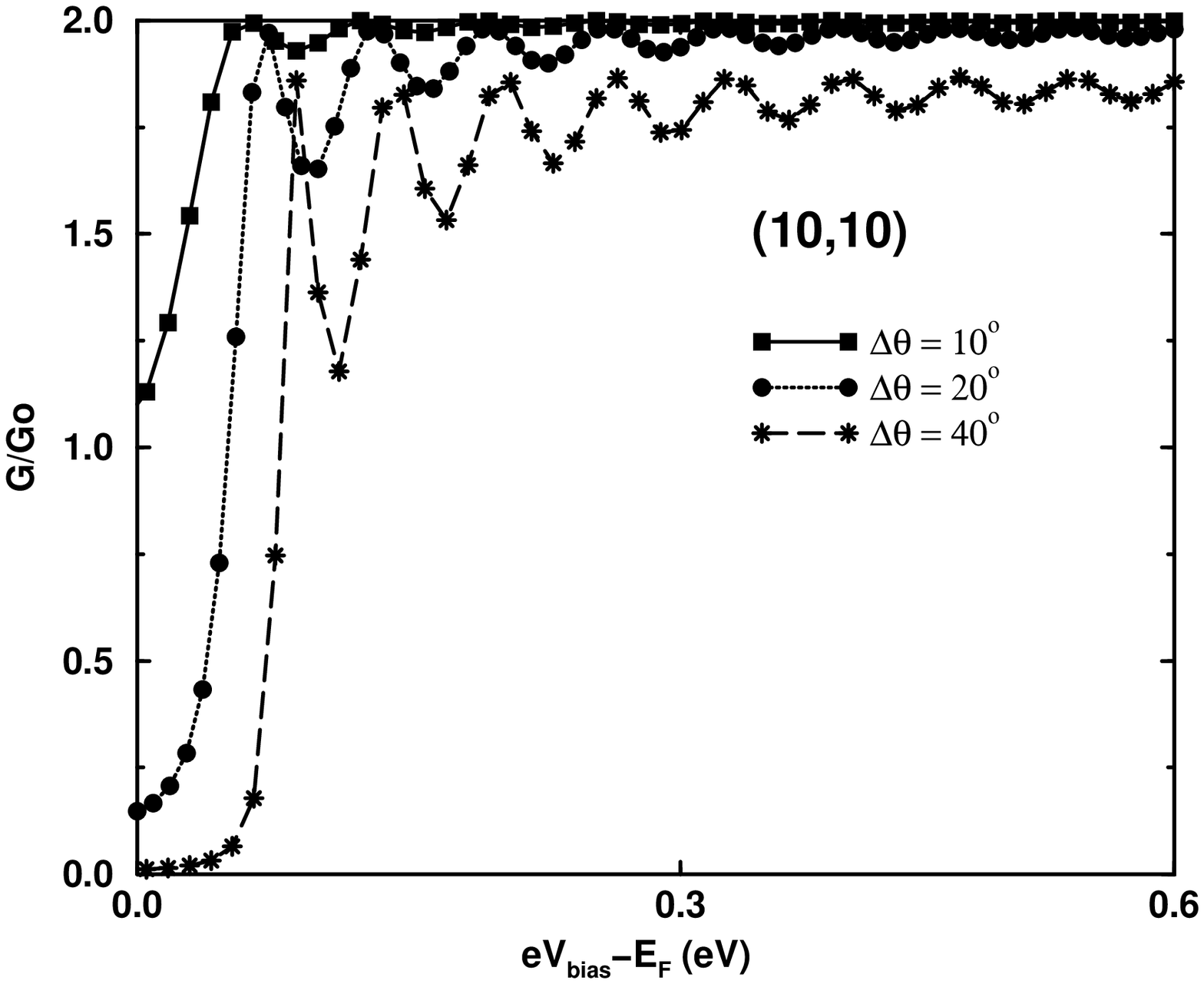}
    \end{minipage}
      \begin{minipage}[c]{0.5\textwidth}
         \hspace*{0.04\columnwidth}
         {\raisebox{0.56\columnwidth}{\Large\bf\textsf {(b)}}}
         \hspace*{-0.09\columnwidth}
         \epsfxsize=0.8\columnwidth
         \epsffile{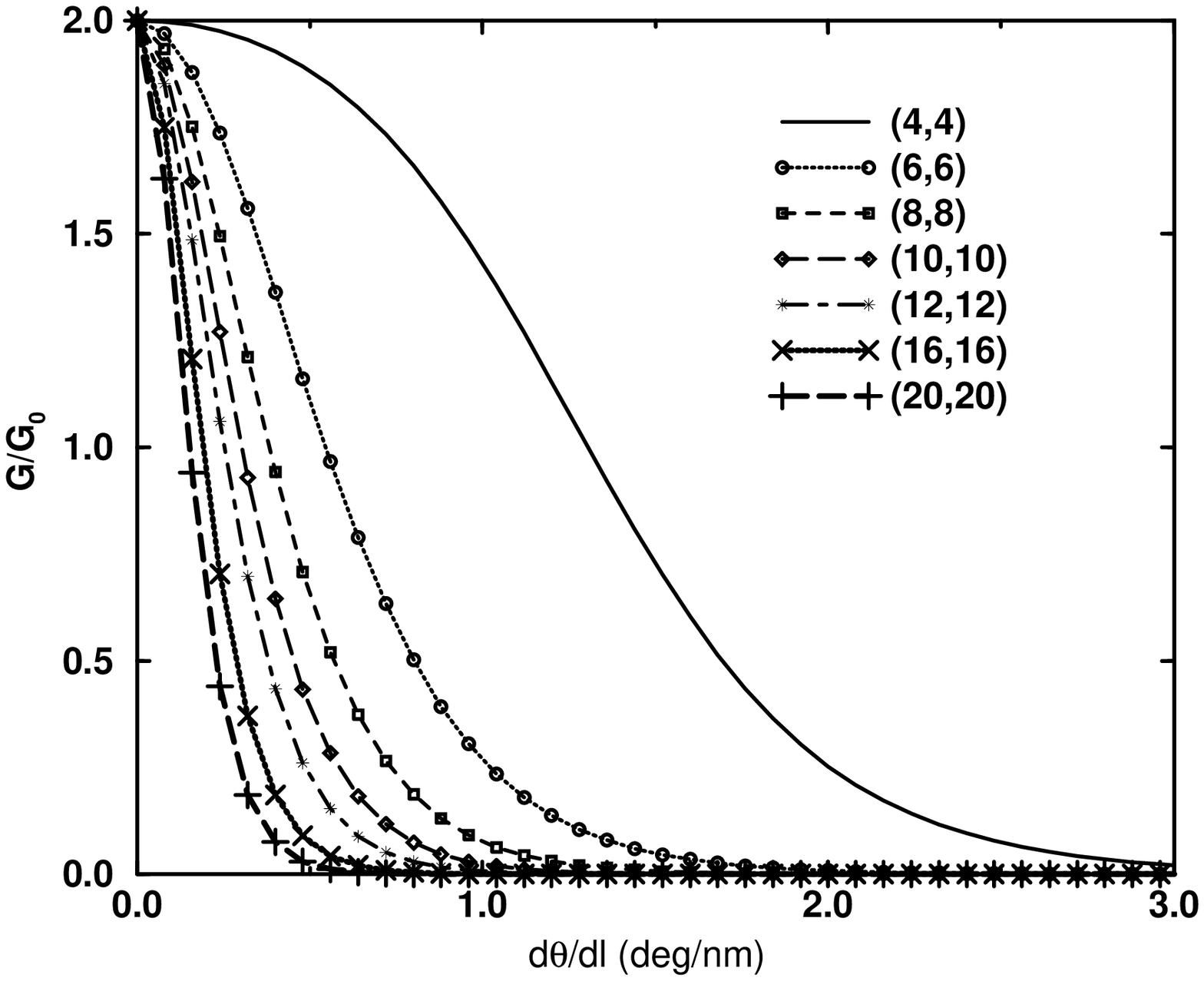}
         \end{minipage}
     \vspace*{0.05\columnwidth}

\caption{ (a) Details of the differential conductance $G$ of a
$(10,10)$ carbon nanotube subject to a finite twiston, as a
function of the applied bias voltage $V_{\rm bias}$. (b)
Differential conductance of $(n,n)$ nanotubes at zero bias as a
function of the local twist in finite twistons. In both figures,
the twistons $\Delta\theta$ extend only across a finite segment of
100 unit cells in the axial direction. } \label{Fig2}
\end{figure}

Details of conductance changes due to localized twistons, such as
those of Fig.~\ref{Fig1}(a), are shown for small bias voltages in
Fig.~\ref{Fig2}(a). In the present case, we subjected a straight
$(10,10)$ nanotube to  finite twists ${\Delta}\theta$ extending
over $100$ unit cells. We find that the  conductance pseudo-gap
associated with finite twistons is accompanied by conductance
oscillations at small bias voltages. A tube subject to an
infinitely long twist, on the other hand, possesses a real gap
with no such oscillations. The zero-bias conductance as a function
of the twist angle for finite twistons extending over $100$ unit
cells is presented in Fig.~\ref{Fig2}(b) for a range of tube
sizes. This type of conductance behaviour is reminiscent of that
associated with tunneling through a potential barrier, where the
twist-induced gap is analogous to the height of the barrier. These
results, when combined, clearly demonstrate that finite twistons
yield non-linear $I-V$ characteristics.

It must be stressed that the degree of twist required to open a
pseudo-gap is small. For example in the finite $(10,10)$ carbon
nanotube of Fig.~\ref{Fig2}(a), ${\Delta}\theta=20^{\circ}$ over
$100$ unit cells is equivalent to ${d\theta/dl} \alt
1^{\circ}/$~{nm} only. The resulting perturbation to the
Hamiltonian is therefore well within the limits of our model,
where the twist is viewed as a frozen-in defect to the nanotube at
low temperatures \cite{PRLxOM}.

We now consider the effect of a uniform magnetic field on
transport in straight and twisted nanotubes. Axially applied
magnetic fields produce an Aharonov-Bohm effect in carbon
nanotubes \cite{AB59,Tian94,Bachtold99}, which in the clean limit
arises from the opening and closing of a band gap at the Fermi
level. In the presence of a twiston, we now demonstrate that the
reverse effect can occur, namely that an axial magnetic field can
remove the twiston-induced conductance gap, resulting in a
positive magneto-conductance.
%%%%%%%%%%%%%%%%%%%CHANGES%%%%%%%%%%%%%%%%%%%%%%%%
Our results show that a magnetic field may restore the zero-bias
conductance of twisted armchair nanotubes from an essentially
vanishing value to near half of the initial zero-twist value. To
understand this behaviour, consider the simplest model of a
twisted nanotube, depicted in the inset of Fig.~\ref{Fig1}(a). Let
us assume that a single $\pi$ orbital per carbon atom is needed to
describe the conductance changes near $E_{\rm F}$. In CNT's the k
vectors perpendicular to the tube surface are quantised and from
\[\Delta({\mathbf k})=|\gamma_0+\gamma_1e^{i\mathbf{k{\cdot}r_1}}+
\gamma_2e^{i\mathbf{k{\cdot}r_2}}|\] in the untwisted armchair CNT
the tight-binding dispersion relation can be calculated to give
\[E_{1D}(k) =\pm \gamma \left(1 + 4
\cos(\frac{q\pi}{n})\cos(\frac{ka}{2})+4\cos^2(\frac{ka}{2})\right)
^{\frac{1}{2}}.\] The structural twist is then modeled by changing
the relative strength of the hopping integral $\gamma_i$ along the
neighbor vector ${\mathbf r_i}$ with respect to the untwisted
reference value $\gamma_0=1$. Upon exposing a twisted tube to a
magnetic field we multiply by a phase factor \cite{Tian94} and
find that the energy gap between the top of the valence and bottom
of the conduction band of a twisted tube changes to
%%%%%%%%%%%%%%%%%%%%%%%%%%%%%%%%%%%%%%%%%%%%%%%%%%%%
\begin{equation}
\Delta({\mathbf k})=|1+\gamma_1e^{i\mathbf{k{\cdot}r_1}}e^{i\phi}+
\gamma_2e^{i\mathbf{k{\cdot}r_2}}e^{i\phi}| \;. \label{Eq1}
\end{equation}
The phase $\phi=2\pi{\Phi}/{\Phi_0}$ depends on the ratio between
the magnetic flux $\Phi$ trapped in the tube and the fundamental
unit of flux $\Phi_0=h/e$. Note that the fundamental unit of flux
used in this paper is twice the flux quantum, $\Phi_0=h/2e$. The
fundamental gap $\Delta=\Delta({\mathbf k_0})$, occurring at
${\mathbf k_0}$, is the minimum value found in the Brillouin zone.
We also note that $\Delta=0$ for an undistorted tube in zero
field.

In view of the twisted tube morphology defined in
Fig.~\ref{Fig1}(a), it is convenient to introduce the quantity
$\alpha=\mathbf{k{\cdot}r_1}=-\mathbf{k{\cdot}r_2}$. From
Eq.~(\ref{Eq1}) we find that $\Delta({\mathbf k})=0$ only if
$(\sin\phi/\sin\alpha)=(\gamma_1-\gamma_2)$ and
$(\cos\phi/\cos\alpha)=(-\gamma_1-\gamma_2)$. Combining these
equations, we find that the fundamental gap closes if
\begin{equation}
\cos{2\phi}=\frac{(\gamma_1^2+\gamma_2^2)-(\gamma_1^2-\gamma_2^2)^2}
{2\gamma_1\gamma_2} \label{Eq2}
\end{equation}
and
\begin{equation}
\cos{2\alpha}=\frac{1-\gamma_1^2-\gamma_2^2}{2\gamma_1\gamma_2}
\;. \label{Eq3}
\end{equation}
In other words, according to Eq.~(\ref{Eq2}), the twist-induced
fundamental gap closes again, once the $(n,n)$ nanotube is exposed
to a magnetic field $B$
\begin{equation}
B = B_c \ \arccos\left(
\frac{(\gamma_1^2+\gamma_2^2)-(\gamma_1^2-\gamma_2^2)^2}{2\gamma_1\gamma_2}
\right) \label{Eq4}
\end{equation}
along its axis, where $B_c=2.28{\times}10^4~$T$/n^2$. According to
Eq.~(\ref{Eq3}), the longitudinal wavevector ${\mathbf k_0}$, at
which the fundamental gap vanishes, must satisfy the condition
\begin{equation}
{\mathbf k}_0{\cdot}{\mathbf r}_1 = \frac{1}{2}\ \arccos\left(
\frac{1-\gamma_1^2-\gamma_2^2}{2\gamma_1\gamma_2} \right) \;.
\label{Eq5}
\end{equation}

The above heuristic model also demonstrates why the exact result
yields ${G}{\approx}{G_0}$ for the zero bias conductance in
presence of a non-zero twist and flux, rather than the value
${G}=2{G_0}$ for a twist-free tube in zero field. In the latter
case, the two open scattering channels at $E_{\rm F}$ correspond
to values of $\alpha$ (or wavevector ${\mathbf k}$) of opposite
sign. For a given nonzero flux $\Phi$ and a corresponding sign of
the phase $\phi$, $(\sin\phi/\sin\alpha)=(\gamma_1-\gamma_2)$ can
be satisfied by only one of these channels.

In summary, we have analyzed for the first time the scattering
properties of finite-size twistons and shown that these introduce
non-linear $I-V$ characteristics associated with the opening of a
pseudo-gap at $E_{\rm F}$, thus effectively quenching the nanotube
conductance. This conductance gap can be closed in an axial
magnetic field, leading to a giant positive magneto-conductance.
At small bias voltages, the conductance of a twisted tube in
nonzero field can reach up to half the ballistic conductance value
of a straight tube in zero field.

YKK and DT acknowledge financial support by the Office of Naval
Research and DARPA under Grant Number N00014-99-1-0252.

%---------------------------------------------------------------------

\end{document}